\documentclass{mn2e}

\usepackage{graphicx,amsmath,amssymb,subfigure}

\def\simgt{\mathrel{\lower0.6ex\hbox{$\buildrel {\textstyle >}
 \over {\scriptstyle \sim}$}}}
\def\simlt{\mathrel{\lower0.6ex\hbox{$\buildrel {\textstyle <}
 \over {\scriptstyle \sim}$}}}
\newcommand{\etal}{{\em et al.}}                   

\newcommand{\ang}{\mbox{$\rm \AA$}}

\begin{document}

\title[Different star formation histories in radio galaxies]
{Evidence of different star formation histories for high- and low-luminosity radio galaxies}

\author[P.D.~Herbert \etal]
{Peter D. Herbert$^{1}$\thanks{Email: p.d.herbert@herts.ac.uk}, Matt J. Jarvis$^{1}$, Chris J. Willott$^{2}$, Ross J. McLure$^{3}$, \and
Ewan Mitchell$^{4}$, Steve Rawlings$^{4}$, Gary J. Hill$^{5}$ and James S. Dunlop$^{3}$ \\
\footnotesize
$^{1}$Centre for Astrophysics Research, Science \& Technology Research Institute,
       University of Hertfordshire, Hatfield, AL10 9AB, UK \\
$^{2}$ Herzberg Institute of Astrophysics, National Research Council,
       5071 West Saanich Rd, Victoria, BC V9E 2E7, Canada\\
$^{3}$SUPA\thanks{Scottish Universities Physics Alliance} Institute
      for Astronomy, University of Edinburgh, Royal Observatory,
      Edinburgh EH9 3HJ\\ 
$^{4}$University of Oxford, Astrophysics, Department of Physics, Keble Road,
       Oxford, OX1 3RH, UK \\
$^{5}$McDonald Observatory, University of Texas at Austin,
       1 University Station C1402, Austin, TX 78712-1083, USA \\
}

\maketitle

\begin{abstract}
We present the results of our investigation into the
stellar populations of 24 radio galaxies at $z \simeq 0.5$
drawn from four complete, low-frequency selected radio
surveys. We use the strength of the 4000\ang\ break as an
indicator of recent star formation, and compare
this with radio luminosity, optical spectral classification
and morphological classification. We find evidence of
different star formation histories for high- and
low-luminosity radio sources; our group of low radio luminosity
sources (typically FRI-type sources) has systematically
older stellar populations than the higher radio luminosity
group. Our sample is also fairly well divided by optical spectral
classification. We find that galaxies classified as having
low excitation spectra (LEGs) possess older stellar populations
than high excitation line objects (HEGs), with the HEGs
showing evidence for recent star formation. We also investigate
the link between radio morphology,
as used by  Owen \& Laing (1989), and the stellar populations.
We find that there is a preference for the ``fat-double'' sources
to have older stellar populations than the ``classical double''
sources, although this is also linked to these sources lying
predominantly in the LEG and HEG categories respectively.
These results are consistent with the hypothesis that HEGs are
powered by accretion of cold gas, which could be supplied,
for example, by recent mergers, secular instabilities,
or filamentary cold flows. These processes
could also trigger star formation in the
host galaxy. The host galaxies of
the LEGs do not show evidence for recent star formation
and an influx of cold gas,
and are consistent with being powered by the accretion of the
hot phase of the inter-stellar medium.
\end{abstract}


\section{Introduction}\label{intro}

There is increasing evidence that active galactic nuclei (AGNs) have
an important role to play in the formation and evolution of
galaxies via AGN-driven feedback.
Various relationships exist between the central black hole of an AGN
and the stellar population of the galaxy: for example, the relation
between black hole mass and stellar velocity dispersion
($M_{BH}$-$\sigma$) (Gebhardt et al. 2000; Ferrarese \&
Merritt 2000; Merritt \& Ferrarese 2001; G\"ultekin et al. 2009)
and the relation between black hole mass and bulge luminosity
(e.g. McLure \& Dunlop 2001; McLure \& Dunlop 2002; G\"ultekin
et al. 2009). The latter may also be expressed as the
Magorrian relation between black hole mass and bulge stellar
mass (Magorrian et al. 1998). These tight correlations
provide compelling evidence for links between galaxy formation
and evolution and the growth of black holes through AGN (accretion) activity.
One suggested mechanism is that of quasar outflows limiting
black hole masses, dependent on the depth of the potential wells
of dark matter halos (Silk \& Rees 1998). 

The current models of galaxy formation and evolution also
invoke AGN-driven feedback to halt the overproduction of stars in the most
massive galaxies. For example, Croton et al. (2006) include AGN-driven feedback
phenomena in their semi-analytic model, separated into ``quasar-mode''
(associated with the efficient accretion of cold gas) and ``radio-mode''
(associated with the less efficient accretion of warm gas) feedback. Sijacki
et al. (2007) also include AGN-driven feedback in their full hydrodynamical model,
once again separated into two modes: the ``quasar regime'' (corresponding
to central black holes with high accretion rates) and mechanical feedback
(corresponding to central black holes with low accretion rates).
Studying the host galaxies of radio-loud AGN should therefore enable
us to place important constraints on the validity of the
feedback mechanisms in the models.

One method of investigating the cold- and hot-mode accretion
in AGNs is with radio samples. Selecting AGNs on the basis of
their low-frequency radio emission enables the  selection of
type-2, or obscured AGNs, in the same way as radio-loud quasars.
This is because the radio waves are unaffected by dust obscuration,
and the radio emission from the extended lobes
is optically thin and therefore orientation independent. Furthermore,
selecting the type-2 radio-loud quasars, or radio galaxies,
allows the study of the stellar populations in the AGN host galaxy,
as the central nuclear emission is obscured by the putative dusty
torus invoked in unified schemes (e.g. Antonucci 1993;
Hill, Goodrich \& DePoy 1996). 
We also note that in the lower power AGNs this optical
faintness may arise not merely from obscuration, but also from an
inherent lack of bright optical-UV continua due to the absence of an
accretion disc (e.g. Chiaberge, Capetti \& Celotti 1999).
The obscuration-related optical faintness
in powerful AGN has enabled many authors to study the evolution of
their host galaxies, with most
work based on the near-infrared ($K-z$) Hubble diagram (Lilly
\& Longair 1982; Jarvis et al. 2001; Willott et al. 2003)
and broad-band colours (e.g. Lilly 1989)
suggesting that the hosts of powerful radio galaxies are comprised
of an old stellar population which forms at high redshift ($z>2$)
and then passively evolves.

Radio galaxies are commonly classified by their radio
morphologies into Fanaroff-Riley type I (FRI) and
Fanaroff-Riley type II (FRII) classes (Fanaroff \&
Riley 1974). An alternative division is that
introduced by Hine \& Longair (1979) based on optical
spectra into high excitation galaxies
(HEGs) and low excitation galaxies (LEGs); this was
further refined by Laing et al. (1994) and
Jackson \& Rawlings (1997) who classified
3CR radio galaxies on this basis (see section~\ref{specclass}).

There is increasing evidence that this 
emission line classification scheme, rather than the
FR-class, has a direct link to the
accretion mode of the AGN.
Hardcastle, Evans \& Croston (2007)
showed that Bondi accretion of the hot phase of the
intergalactic medium (IGM) is sufficient to power
all low excitation radio sources, whereas the HEGs are powered by the
accretion of cold gas typically thought to be driven
towards the central engine during a galaxy merger.
There are, however, other theories for the origin of
the accreted cold gas in these galaxies. For example, in the
hydrodynamical model of Ciotti \& Ostriker (2007), 
gas from dying stars can be recycled by cooling radiatively and
falling to the nuclear region where a small fraction ($\approx 1 \%$
or less) is accreted onto the central black hole.
Alternatively, various authors (Kere\v s et al. 2005; Dekel \& Birnboim 2006;
Ocvirk, Pichon \& Teyssier 2008; Dekel et al. 2009; Brooks et al. 2009; Kere\v s et al. 2009)
have used hydrodynamic simulations to show that cold mode accretion
via streams or filaments of cold gas could be responsible for
the majority of the gas supplied to massive galaxies at high redshifts.
For lower mass galaxies this cold mode accretion could be the dominant
source of cold gas right up to the present day.


Following on from earlier work by Lilly \& Longair (1984),
Baldi \& Capetti (2008) studied a sample of nearby
3CR radio galaxies and compared optical and UV
images in order to detect evidence of recent star
formation.  They found evidence of recent star
formation in the HEGs in their sample, but not in
the LEGs. They suggest that the HEGs have undergone
recent major mergers which both trigger star
formation and power the AGN by providing a supply
of cold gas for accretion. The LEGs in their
picture, which supports the one put forward by
Hardcastle et al. (2007), have experienced no such
mergers.  In this case no bursts of star
formation are triggered and the AGN are
powered by accretion of the hot interstellar
medium (ISM).

In complementary work, Emonts et al. (2008) find no evidence
for large-scale H{\sc i} structures in the host galaxies of
FRI-type sources, but find that the host galaxies of
most of the FRII-type sources in their sample contain
significant amounts of H{\sc i}. They suggest that this
dichotomy is a result of different formation histories
for FRI and FRII-type radio sources. This is again consistent
with the picture outlined above in which high- and  low-luminosity
radio galaxies have undergone different star formation histories
and this also influences the mode of accretion on to the central
supermassive black hole.

In recent work using the vast data set available from the Sloan
Digital Sky Survey, Kauffmann, Heckman \& Best (2008) study a
sample of radio-loud AGNs with emission lines. They find a strong
correlation between the presence of
emission lines and the presence of a young stellar
population, consistent with the findings of
Baldi \& Capetti (2008). They also find a correlation
between the age of the stellar population and the
radio luminosity normalised by the black hole mass.
They note that strong optical AGNs have a significantly
enhanced probability of hosting radio jets, and
thus the radio and optical phenomena are not
independent. However, although the SDSS is a
wide-area survey, it does not probe the necessary
volume to obtain a significant sample of the more luminous FRII-type or HEG
populations. Therefore, in order to study these
effects further, we need to use a complete sample of
radio sources selected by other methods.

In this paper we present the results of our
study into the star formation histories (SFHs)
of 24 radio galaxies at $0.4 < z < 0.6$. 
We compare our 4000\ang\ break indices with
radio luminosity, spectral classification
and morphological classification.
We begin in section \ref{thesample} of the paper by
describing our sample, our observations and our data
reduction. Section \ref{analysis} describes our
4000\ang\ break strength measurements and 
our spectral and morphological classifications. We present
our findings in section \ref{discussion} before
concluding in section \ref{conclusions}.
Throughout
the paper we assume a standard cosmology in which
$H_{0}$ = 70 km s$^{-1}$, $\Omega_{M}$ = 0.3 and
$\Omega_{\Lambda}$ = 0.7.

\section{The Sample}\label{thesample}

Our full 41-object $z \simeq 0.5$ radio-galaxy sample
(the ZP5 sample)
consists of all of the radio galaxies without broad
optical emission lines in the redshift
interval $0.4 < z < 0.6$ from four complete, low-frequency
selected radio surveys; 3CRR (Laing, Riley \& Longair 1983), 6CE
(Eales et al. 1997; Rawlings, Eales \& Lacy 2001), 7CRS (Lacy et al.
1999; Willott et al. 2003) and TexOx-1000 (Hill \& Rawlings 2003;
Vardoulaki et al. 2009).
The objects from this last survey probe towards the SDSS
regime (Kauffmann et al. 2008),
although SDSS galaxies still have a lower luminosity in the radio.
Full details of the sample, along with the motivations for this
choice of sample, can be found in McLure et al. (2004).

Using the 4.2-metre William Herschel Telescope (WHT) and the
8.1-metre Gemini North telescope we have obtained
optical spectra for a subsample comprising 24 of the
objects from the full ZP5 sample. Details of the subsample
can be found in Table \ref{table:sample}. Objects
observed using the WHT were chosen to have $z < 0.5$ so that
satisfactory signal to noise ratios could be obtained
without excessively long integration times. This constraint
aside, the objects in our subsample were chosen at random
from the objects in the full sample that were visible
on the dates of observation. Details of the observations and
data reduction can be found in Herbert et al. (in preparation).

\begin{table*}
\centering
\caption{\label{table:sample}Our $z \simeq 0.5$ subsample.  Column 1 lists the
radio galaxy names and columns 2 and 3 list the J2000 source
coordinates.  Column 4 lists the object redshifts, column 5 the
logarithm of the 151-MHz luminosities in units of W~Hz$^{-1}$~sr$^{-1}$, and
column 6 the telescope used for the spectroscopic observations.
Column 7 lists the calculated D$_{n}$(4000) indices, and column
8 our emission line classification (HEG or LEG).
Column 9 lists our morphological
classification into Classical Doubles (CD),
Jetted sources (J) and Fat Doubles (FD).
TOOT1648+5040 was too compact for us to classify.}
\begin{tabular*}{16.4cm}{lcccccccc}
\hline
Source & RA & Dec & z & L$_{151}$ & Telescope & D$_{n}$(4000) & Spec. Class & Morph. Class \\
\hline
3C16 & 00 37 45.39 & +13 20 09.6 & 0.405 & 26.82 & WHT & $1.54 \pm 0.10$ & HEG & CD/FD \\
3C19 & 00 40 55.01 & +33 10 07.3 & 0.482 & 26.96 & WHT & $1.59 \pm 0.04$ & LEG & CD \\
3C46 & 01 35 28.47 & +37 54 05.7 & 0.437 & 26.84 & WHT & $1.47 \pm 0.04$ & HEG & CD \\
3C172 & 07 02 08.32 & +25 13 53.9 & 0.519 & 27.17 & Gemini & $1.63 \pm 0.05$ & HEG & CD \\
3C200 & 08 27 25.38 & +29 18 45.5 & 0.458 & 26.92 & Gemini & $1.55 \pm 0.03$ & LEG & CD \\
3C244.1 & 10 33 33.97 & +58 14 35.8 & 0.428 & 27.10 & WHT & $1.30 \pm 0.03$ & HEG & CD \\
3C295 & 14 11 20.65 & +52 12 09.0 & 0.464 & 27.68 & Gemini & $1.58 \pm 0.04$ & HEG & CD \\
3C341 & 16 28 04.04 & +27 41 39.3 & 0.448 & 26.88 & WHT & $1.34 \pm 0.04$ & HEG & CD \\
3C427.1 & 21 04 07.07 & +76 33 10.8 & 0.572 & 27.53 & Gemini & $1.52 \pm 0.08$ & LEG & CD \\
3C457 & 23 12 07.57 & +18 45 41.4 & 0.428 & 27.00 & WHT & $1.24 \pm 0.03$ & HEG & CD \\
\\
6C0825+3407 & 08 25 14.59 & +34 07 16.8 & 0.406 & 26.09 & WHT & $1.65 \pm 0.10$ & LEG & FD \\
6C0850+3747 & 08 50 24.77 & +37 47 09.1 & 0.407 & 26.15 & WHT & $1.25 \pm 0.04$ & HEG & CD \\
6C0857+3945 & 08 57 43.56 & +39 45 29.0 & 0.528 & 26.34 & Gemini & $1.46 \pm 0.03$ & HEG & CD \\
6C1303+3756 & 13 03 44.26 & +37 56 15.2 & 0.470 & 26.29 & Gemini & $1.56 \pm 0.01$ & HEG & CD \\
\\
7C0213+3418 & 02 13 28.39 & +34 18 30.6 & 0.465 & 25.66 & WHT & $1.82 \pm 0.06$ & LEG & FD \\
7C0219+3423 & 02 19 37.83 & +34 23 11.2 & 0.595 & 25.98 & Gemini & $1.17 \pm 0.02$ & HEG & FD \\
7C0810+2650 & 08 10 26.10 & +26 50 49.1 & 0.435 & 25.58 & WHT & $1.40 \pm 0.04$ & HEG & CD \\
7C1731+6638 & 17 31 43.84 & +66 38 56.7 & 0.562 & 25.52 & Gemini & $1.21 \pm 0.03$ & HEG & FD \\
7C1806+6831 & 18 06 50.16 & +68 31 41.9 & 0.580 & 26.36 & Gemini & $1.58 \pm 0.02$ & HEG & FD \\
\\
TOOT0009+3523 & 00 09 46.90 & +35 23 45.1 & 0.439 & 24.79 & WHT & $1.68 \pm 0.06$ & LEG & FD \\
TOOT0018+3510 & 00 18 53.93 & +35 10 12.1 & 0.416 & 25.16 & WHT & $1.69 \pm 0.04$ & LEG & J? \\
TOOT1626+4523 & 16 26 48.50 & +45 23 42.6 & 0.458 & 25.03 & WHT & $1.76 \pm 0.05$ & LEG & FD \\
TOOT1630+4534 & 16 30 32.80 & +45 34 26.0 & 0.493 & 25.17 & WHT & $1.74 \pm 0.03$ & LEG & J \\
TOOT1648+5040 & 16 48 26.19 & +50 40 58.0 & 0.420 & 25.12 & WHT & $1.90 \pm 0.07$ & LEG & - \\
\hline
\end{tabular*}
\end{table*}

\section{Analysis}\label{analysis}

\subsection{4000\AA\ Break Strength}\label{d4000}

The strength of the 4000\ang\ break can be used as a measure
of recent star formation, as outlined by Kauffmann et al.
(2003). Younger stellar populations contain hotter stars
which have multiply ionized elements.  This leads to a
decrease in opacity and consequently a smaller 4000\ang\
break. Conversely, the 4000\ang\ break is larger for older
populations.

\begin{figure}
\centering
\includegraphics[width=0.5\textwidth]{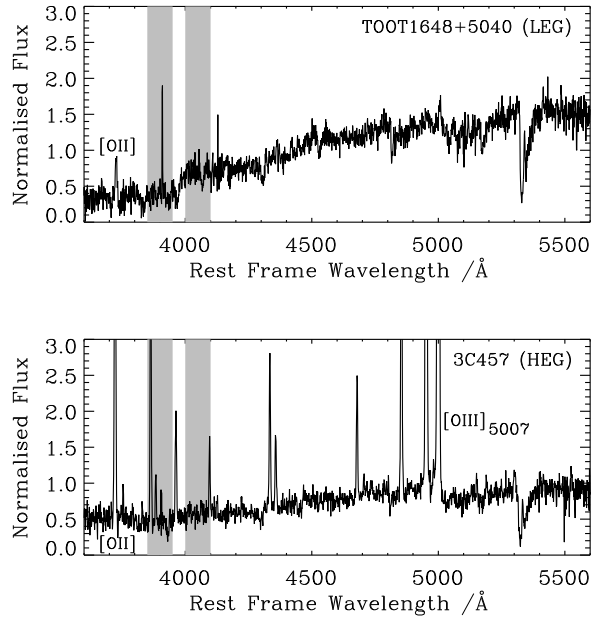}
\caption{\label{spectra}The spectra of two of our
objects: TOOT1648+5040 (a low-excitation radio
galaxy with a large 4000\ang\ break) and 3C457
(a high-excitation radio galaxy with a smaller 4000\ang\
break). The shaded regions indicate the bands used for
the D$_{n}$(4000) calculation (Section \ref{d4000}).}
\end{figure}

In order to measure the strength of the 4000\ang\ break
we adopt the band definitions of Balogh et al. (1999)
as used by Kauffmann et al. (2003);
the break strength [D$_{n}$(4000)] is calculated as the
mean flux in the 4000-4100\ang\ band divided by
the mean flux in the 3850-3950\ang\ band (see
Figure \ref{spectra}). We calculate
D$_{n}$(4000) for each object directly from our spectra,
masking out regions affected by emission lines
from the central radio source.

Our calculated
D$_{n}$(4000) indices can be found in
Table \ref{table:sample}.
Our D$_{n}$(4000) indices lie in the range 1.17-1.90.
From figure 6 of Kauffmann et al. (2003) we note
that the galaxies at the upper end of this range
correspond to old galaxies with no recent star
formation. Those galaxies at the lower end of the range
are consistent with galaxies where a small fraction
of the stellar mass has been formed 0.1-2 Gyr ago.

The D$_{n}$(4000) index can be
contaminated by dust and/or scattered quasar light.
There are thus potential limitations in using
D$_{n}$(4000) in isolation as an indicator
of star formation histories. An additional
indicator is the H$\delta$ index (see
Worthey \& Ottaviani 1997; Kauffmann et al. 2003).
However, we are
unable to measure H$\delta$ for our galaxies
due to the presence of emission lines from the
AGN. We therefore use D$_{n}$(4000) by itself,
noting that there is no evidence for dust in
any of our objects, i.e. they are all adequately
fit by simple stellar populations without dust
obscuration (Herbert et al. in preparation) and include
the Ca H+K lines giving us confidence that the contribution
to the blue light is stellar in origin.
In Section \ref{discussion}
we also show that the D$_{n}$(4000) indices
are unlikely to be affected by scattered quasar
light.

\subsection{Spectral and Morphological Classifications}\label{specclass}

We classify our radio galaxies as high-excitation (HEGs)
or low-excitation (LEGs) according to the classification
scheme of Jackson \& Rawlings (1997). An object is
classified as a LEG if it has an [OIII] rest frame equivalent width
$< 10 \ang$ or an [OII]/[OIII] ratio $> 1$ (or both).
Our classifications can be found in Table
\ref{table:sample}
\footnote{For objects where our wavelength range
does not extend to the [OIII] emission line
we use data from
http://www.science.uottawa.ca/$\sim$cwillott/3crr/3crr.html}. 
Figure \ref{spectra} illustrates the difference in the spectra
between high-excitation and low-excitation objects.


To morphologically classify the radio source we adopt the
classification scheme of Owen \& Laing (1989) and
classify our sources as Classical Doubles (CD), Jetted sources
(J) or Fat Doubles (FD). In terms of their Fanaroff-Riley
classification (Fanaroff \& Riley 1974), Classical Doubles are
always identified with FRII-type sources whilst Fat Doubles are
identified as FRII or FRI/II radio galaxies and Jetted
sources are generally identified with FRI-type sources.
Our classifications can be found in Table
\ref{table:sample} and full details can be found in
Mitchell (2006).

\section{Discussion}\label{discussion}

\subsection{D$_{n}$(4000) versus Radio Luminosity}\label{d4000radlum}

\begin{figure}
\centering
\includegraphics[width=0.5\textwidth]{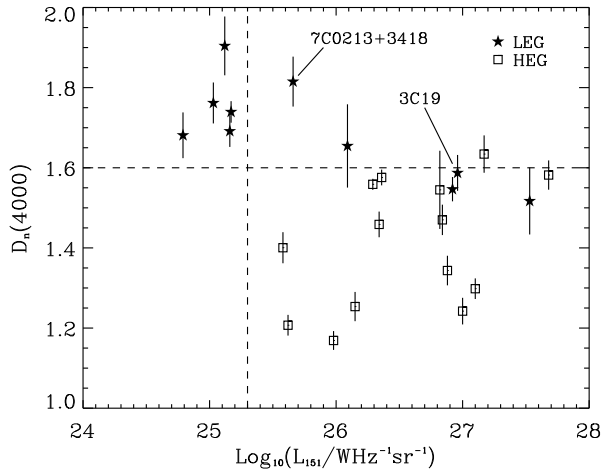}
\caption{\label{heglegplot}D$_{n}$(4000) versus L$_{151}$,
split by spectral classification. LEGs are shown as
filled stars and HEGs as open squares.
The vertical dashed line is included to highlight the division
between the two populations discussed in the text.
The horizontal dashed line highlights the dichotomy in
D$_{n}$(4000) for the two populations.}
\end{figure}

In Figure \ref{heglegplot} we show D$_{n}$(4000) versus
the low-frequency radio luminosity at 151~MHz, L$_{151}$,
for our galaxies, with symbols corresponding to their
spectral classification. Motivated by figure 6 of Kauffmann
et al. (2003), we include a dashed line at D$_{n}$(4000) = 1.6
in order to illustrate differences in star formation. Objects
with D$_{n}$(4000) $<$ 1.6 may have formed a noticeable fraction
(5\% or greater) of their stellar mass in recent star bursts in
the models of Kauffmann et al. (2003).  On the other hand, objects with 
D$_{n}$(4000) $>$ 1.6 show little or no evidence for recent star
formation. This division at D$_{n}$(4000) = 1.6 therefore represents
a conservative division between objects which may have evidence of
recent star formation, and those which do not. We reiterate that
since we are unable to measure
H$\delta$ for our objects we are unable to be more specific
as regards the star formation histories of our galaxies.

Mindful of this note of caution, it is still
readily apparent that the
galaxies form two distinct populations. One population,
with lower radio luminosities (L$_{151} < 10^{25.3}$
W~Hz$^{-1}$~sr$^{-1}$), is composed exclusively
of LEGs. The galaxies in this population have systematically
higher D$_{n}$(4000) indices and thus older stellar populations.
The second population, at higher radio luminosities
(L$_{151} > 10^{25.3}$ W~Hz$^{-1}$~sr$^{-1}$) and
smaller D$_{n}$(4000) indices (younger stellar populations),
consists mainly of HEGs, although a few LEGs are also present
in this population. 6C0825+3407 and 7C0213+3418
could belong to either population (as discussed in
Section \ref{d4000-em}).
We illustrate the division between the two populations
by the vertical dashed line in Figure \ref{heglegplot}. The traditional
division between FRI type and FRII type radio galaxies
falls at L$_{151} = 10^{25.3}$ W~Hz$^{-1}$~sr$^{-1}$ and
motivates our population division at this value.
This is also the radio luminosity at which there is
an apparent divergence in the evolution with redshift,
the higher luminosity radio sources tending to evolve more
strongly than the lower luminosity sources
(Clewley \& Jarvis 2004; Sadler et al. 2007).
Performing a two sided K-S test on the D$_{n}$(4000) indices
for the two populations enables us to reject at a
significance of 99.96\% that the two populations are
drawn from the same distribution.
We also perform a Mann-Whitney-Wilcoxon (MWW) test
on the two populations and reject at a significance
of $>99.9\%$ the null hypothesis that they are drawn
from the same distribution.
We thus find evidence for different star formation
histories for high- and low- luminosity radio galaxies. The
transition occurs around L$_{151} \simeq 10^{25.3}$
W~Hz$^{-1}$~sr$^{-1}$, although it
is not possible to make a clean divison at a single value
of  L$_{151}$.

\subsection{D$_{n}$(4000) versus Emission-line Classification}\label{d4000-em}

\begin{figure}
\centering
\includegraphics[width=0.5\textwidth]{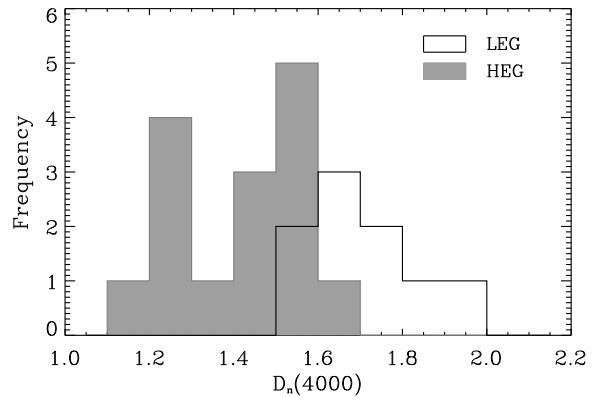}
\caption{\label{histograms} A histogram of the D$_{n}$(4000) of
our objects, with HEGs (shaded) and LEGs (clear) shown
separately.}
\end{figure}

We also split our objects according to their optical
classification (i.e. LEGs and HEGs) and perform another
two sided K-S test on the D$_{n}$(4000) indices.
In this case we reject the null hypothesis that the indices
are drawn from the same distribution at a significance
of 99.4\%. An MWW test in this case rejects the null
hypothesis at a significance of 99.8\%. The difference
in D$_{n}$(4000) between a LEG and a HEG is illustrated in
Figure \ref{spectra} and Figure \ref{histograms} shows clearly that
our HEGs and LEGs are divided into two separate populations on
the basis of their D$_{n}$(4000) indices. We thus find
evidence in support of the work of Kauffmann et al. (2008)
and Baldi \& Capetti (2008). The latter authors
find evidence of recent star formation in HEGs but
not in LEGs. They suggest that HEGs have undergone a recent
major merger that triggered star formation and
also provided the fuel to power the AGN via cold gas accretion.
In their picture,
LEGs on the other hand have had no such recent merger, and
thus are fuelled by the hot ISM and show no
evidence of recent star formation.
However, whilst mergers seem a likely explanation of
the origin of the influx of cold gas in HEGs, we note that
there are alternative explanations of this influx, as discussed
in Section \ref{intro}. We also
note that our population that shows evidence of
more recent star formation contains three (or possibly four)
LEGs. Thus a spectral line classification system may not be
a clean method to study the history of the influx of cold gas into
AGN hosts -- rather the star formation history measured by studying the host galaxies may
offer a better indication of the past influx of cold gas by
mergers or other processes.

However, it is possible that even when the principal accretion
mechanism is through the hot-mode, minor mergers and interactions
could stimulate some star-formation activity which would result
in smaller D$_{n}$(4000) values. Furthermore, the timescale
on which a merger influences the AGN fuelling could be different
from the timescale on which it triggers star formation which
would also cause mixing of the populations.
It is also interesting to note that the LEGs still have amongst
the oldest stellar populations in the higher radio luminosity
population.
We also observe that some of the LEGs belonging to the
higher radio luminosity population possess HEG-like features.
3C19 (labelled in Figure \ref{heglegplot}) would be classed as
a HEG on the basis of its [OIII] rest frame equivalent width,
but is classified as a LEG due to its [OII]/[OIII] ratio.
Conversely 7C0213+3418 (also labelled in Figure \ref{heglegplot})
would be a HEG on the basis of its [OII]/[OIII] ratio, but is
classified as a LEG because of its [OIII] rest frame equivalent width.

An important question is whether the dichotomy observed in
D$_{n}$(4000) is primarily related to the radio luminosity
(Section \ref{d4000radlum}) or to the excitation state (this section).
Given our current data we are unable to provide a conclusive answer
to this question. However, given that the excitation state
appears to be a better indicator of the accretion rate than the
radio luminosity (Hardcastle et al. 2007), and thus is more closely linked to the state of
the gas, we suggest that it is the excitation state, rather than
the radio emission, which is primarily related to the strength of the 
4000\AA\ break (and thus the star formation history of the galaxy).

\begin{figure}
\centering
\includegraphics[width=0.5\textwidth]{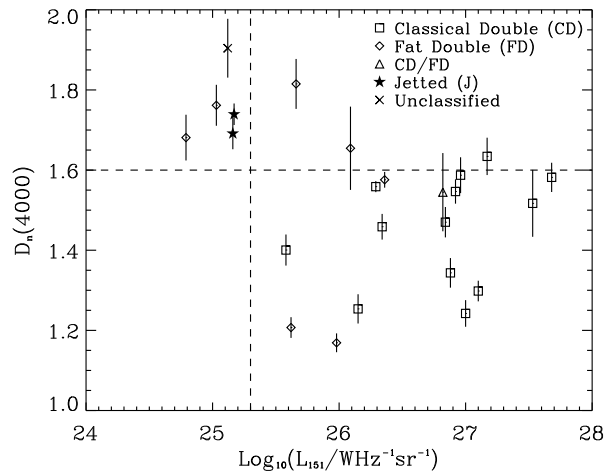}
\caption{\label{morphologyplot}D$_{n}$(4000) versus L$_{151}$,
split by radio morphology. Classical Doubles are represented
by open squares and Fat Doubles by open diamonds. `Classical
Double/Fat Doubles' are shown as open triangles.
Jetted sources are represented by filled stars and
unclassified objects are marked with a cross.
The dashed lines are as in Figure \ref{heglegplot}.}
\end{figure}

In Figure \ref{morphologyplot} we again show D$_{n}$(4000)
versus L$_{151}$, this time with symbols corresponding
to the radio morphology of the galaxy. We find that our
lower radio luminosity population is
composed mainly of Fat Doubles and our higher radio luminosity
population
contains predominantly Classical Doubles.
However, there
is a great deal of mixing of the radio morphology types, and
we find no clear difference between the two populations on the basis
of radio morphology \textbf{alone} due to the strong relation between radio morphology
classification and the high- and low-excitation populations
(Hine \& Longair 1979).

\subsection{Contamination by Scattered Quasar Light?}

Scattered quasar light has, in the past, been posited as an
explanation for the alignment effect in radio galaxies (e.g.
Tadhunter et al. 1992, Cimatti et al. 1993) and could possibly
lead to uncertainties in the inferred properties of quasar
host galaxies (e.g. Young et al. 2009). 
In this paper we use the strength of the 4000\ang\ break as an
indicator of recent star formation. Care must therefore
be taken to consider whether the dilution in
the 4000\ang\ break arises as a result of a
young stellar population or from other sources.
Lilly \& Longair (1984) and Lilly, Longair \&
Allington-Smith (1985) used the 3CR and `1-Jansky'
samples of radio galaxies to establish the trend
for bluer galaxies to have stronger [OII]3727
line emission (from the AGN). More recent work (e.g. Tadhunter,
Dickson \& Shaw 1996; Aretxaga et al. 2001; Tadhunter et al. 2002;
Holt et al. 2007) has emphasized the contribution
of both young stellar populations and
AGN-related components (e.g. nebular continuum,
scattered and direct quasar light, emission lines)
to the optical and UV continua in powerful radio galaxies.

As found by Lilly et al. (1985), we would expect the
contamination by scattered quasar light to be highly correlated
with the ionizing power of the central engine and thus the
emission-line luminosity, assuming emission-line luminosity is
a good proxy for ionizing power (see e.g. Rawlings \& Saunders 1991).
However, we find no evidence for a correlation between the [OII]
luminosity and D$_{n}$(4000) with a Spearman Rank test giving only a
25 per cent probability that the relation deviates from the null
hypothesis of no correlation. 

Furthermore, for scattered quasar light to have a large effect on
our results would require a very strong dependence on wavelength,
i.e. over the range where we calculate D$_{n}$(4000). Using the
most pronounced examples of scattered quasar light in the
literature, i.e. in broad-line radio galaxies (see e.g. Tran et
al. 1998) we find, by using the total polarized emission spectrum
from these sources, that the spectrum of the scattered light
gives a value of D$_{n}$(4000)$\approx 0.9$. 
Assuming that at 3900\ang\ up to 10 per cent of the total emission in our
radio galaxies could be contributed by the scattered nuclear
emission (e.g. figure 3 of Tran et al. 1998)
we calculate that this would reduce our D$_{n}$(4000)
by a maximum value of 0.1. Given that none of our radio galaxy
spectra show any evidence for any scattered contribution,
have pronounced absorption-line spectra and are all well fit
by a simple stellar population, along with the lack of a
correlation between [OII] emission-line luminosity and
D$_{n}$(4000), we are confident that the dichotomy in D$_{n}$(4000)
between the HEGs and LEGs is due to different star-formation
histories rather than scattered quasar light.




\section{Conclusions}\label{conclusions}

In this paper we have used deep spectroscopic observations to
determine how the age of the stellar populations in the host
galaxies of powerful radio sources is related to the structure
of the radio emission and the ionizing power of the AGN.

\begin{itemize}

\item We have shown that our sample of $z \simeq 0.5$ radio galaxies
forms two distinct populations in the D$_{n}$(4000) - L$_{151}$ plane.
The population at lower radio luminosity
is composed entirely of LEGs, the one at higher radio luminosity
predominantly of HEGs (although a few LEGs fall within this population).

\item Our lower radio luminosity population has systematically
higher D$_{n}$(4000) indices than the higher radio luminosity
population. We find that this is most likely due to the presence
of a younger stellar population in the higher radio luminosity
(or HEG) population after considering the possibility that
scattered quasar light may significantly alter the D$_{n}$(4000).
We find that scattered quasar light could only decrease D$_{n}$(4000)
by a maximum value of 0.1, which is not sufficient to explain the
dichotomy that we observe. However, we are unable to
measure the H$\delta$ for our objects, which does leave some
uncertainty in determining the actual extent of the star formation
in our galaxies.

\item We find that there is a preference for ``fat-double'' sources
to have older stellar populations than the ``classical double''
sources. However, this may simply reflect the underlying links
between morphology and excitation state or radio morphology,
which appear to be more strongly correlated with D$_{n}$(4000) than
is the morphology.

\item Our results are consistent with the hypothesis that HEGs
are powered by the accretion of cold gas, the influx of
which could be due to mergers -- which would also trigger star
formation -- or other processes, such as cooled, recycled gas from
dying stars or cold mode accretion. On the other hand, we
suggest that LEGs are powered by the accretion of the hot phase of
the inter-stellar medium (as they have undergone no recent influx
of cold gas).


\end{itemize}

The results presented here show that deep optical spectroscopy
can provide important information on the host galaxy properties
of moderate redshift, powerful radio galaxies. However, this also
needs to be linked to other indicators of star formation and merger
activity to obtain a comprehensive and consistent explanation for
the observed differences between AGN and their corresponding effect
on the evolution of massive galaxies.

In future work (Herbert et al. in preparation) we will use this sample
to investigate where these radio sources fall on the Fundamental
Plane and how they relate to the general massive galaxy population.
Furthermore, we also have deep multi-band imaging data to assess
the environmental richness of the full $z \simeq 0.5$ sample which
will enable us to link how the power of the AGN, its host galaxy mass
and star formation history are related to the environmental
density of the AGN.

\section*{ACKNOWLEDGEMENTS} 
{\footnotesize}
We thank Martin Hardcastle for useful discussions and
the anonymous referee for careful reading and for suggestions
which have improved the manuscript.
PDH thanks the UK STFC for a studentship. MJJ ackowledges the
support of an RCUK fellowship. RJM and JSD acknowledge the support
of the Royal Society through a University Research Fellowship and
a Wolfson Research Merit award respectively.
The William Herschel Telescope is operated on the
island of La Palma by the Isaac Newton Group in the Spanish
Observatorio del Roque de los Muchachos of the Instituto de Astrofisica
de Canarias.
Based on observations obtained at the Gemini Observatory
(programs GN-2008B-Q-103 \& GN-2009A-Q-105),
which is operated by the Association of Universities for
Research in Astronomy, Inc., under a cooperative agreement
with the NSF on behalf of the Gemini partnership: the
National Science Foundation (United States), the Science
and Technology Facilities Council (United Kingdom), the
National Research Council (Canada), CONICYT (Chile), the
Australian Research Council (Australia), Minist\'erio da
Ci\^encia e Tecnologia (Brazil) and Ministerio de Ciencia,
Tecnolog\'ia e Innovaci\'on Productiva (Argentina).

{}

\end{document}